\documentclass[twocolumn,showpacs,preprintnumbers,amsmath,amssymb]{revtex4}

\usepackage{graphicx}
\usepackage{dcolumn}
\usepackage{bm}

\begin{document}

\preprint{APS/123-QED}

\title{Elementary excitations of ${\bm S}$\,{=}\,1/2 one-dimensional antiferromagnet KCuGaF$_6$\\in magnetic field and quantum sine-Gordon model}

\author{Izumi Umegaki$^1$}
  \email{umegaki@lee.phys.titech.ac.jp.}
\author{Hidekazu Tanaka$^1$}
  \email{tanaka@lee.phys.titech.ac.jp.}
\author{Toshio Ono$^1$}
\author{Hidehiro Uekusa$^2$}  
\author{Hiroyuki Nojiri$^3$} 
  \email{nojiri@imr.tohoku-u.ac.jp.}
\affiliation{$^1$Department of Physics, Tokyo Institute of Technology, Meguro-ku, Tokyo 152-8551, Japan\\
$^2$Department of Chemistry and Materials Science, Tokyo Institute of Technology, Meguro-ku, Tokyo 152-8551, Japan\\
$^3$Institute for Material Research, Tohoku University, Aoba-ku, Sendai 980-8577, Japan}

\date{\today}

\begin{abstract}
Elementary excitations of the $S\,{=}\,1/2$ one-dimensional Heisenberg antiferromagnet KCuGaF$_6$ with exchange constant $J/k_{\rm B}\,{=}\,103$ K were investigated by high-frequency ESR measurements combined with a pulsed high magnetic field. When an external magnetic field $H$ is applied in KCuGaF$_6$, a staggered magnetic field $h$ is induced perpendicular to $H$ owing to the staggered $\bm g$ tensor and the Dzyaloshinsky-Moriya (DM) interaction with an alternating $\bm D$ vector. Consequently, KCuGaF$_6$ in a magnetic field is represented by the quantum sine-Gordon (SG) model. We observed many resonance modes including a soliton resonance, breathers, interbreather transitions and two-breather excitation. Their resonance conditions are beautifully described by the quantum SG field theory with one adjustable parameter $c_{\rm s}\,{=}\,h/H$. To investigate the relationship between the Curie term due to the DM interaction and the proportional coefficient $c_{\rm s}$, magnetic susceptibility measurements were also performed varying the external field direction. 
\end{abstract}

\maketitle


\section{Introduction}
Elementary excitations of the antiferromagnetic Heisenberg chain (AFHC) are different from those calculated from the linear spin wave theory\,\cite{Anderson,Kubo} and are complex because of the strong quantum fluctuation characteristic of one dimension. For the $S\,{=}\,1/2$ uniform AFHC, the elementary excitations are $S_z\,{=}\,1/2$ excitations called spinons and their nature is well understood with the help of an exact solution\,\cite{dCP,Ishimura,Faddeev} and accurate analytical and numerical calculations\,\cite{Yamada,Pytte,Muller}. The theoretical results for zero magnetic field were verified by neutron inelastic scattering experiments\,\cite{Endoh,Nagler}. The most prominent difference between the linear spin wave theory and accurate results\,\cite{Pytte,Ishimura} is observed in the dispersion relation under an external magnetic field $H$\,\cite{Heilmann}. Accurate analyses demonstrated that the gapless excitations occur at incommensurate wave vectors $q\,{=}\,{\pm}\,2{\pi}m(H)$ and ${\pi}\,{\pm}\,2{\pi}m(H)$ in addition to at $q\,{=}\,0$ and ${\pi}$, where $m(H)$ is the magnetization per site in the unit of $g{\mu}_{\rm B}$. 

An attempt to observe such incommensurate gapless excitation was made by Dender {\it et al.}\,\cite{Dender}, who performed neutron inelastic scattering experiments and specific heat measurements in Cu(C$_6$H$_5$COO)$_2$$\cdot$3H$_2$O called copper benzoate under magnetic fields. However, they observed an unexpected excitation gap ${\Delta}(H)$ proportional to $H^{2/3}$. Copper benzoate is known to be an $S\,{=}\,1/2$ antiferromagnet with good one-dimensionality\,\cite{Date1,Date2,Takeda}. Oshikawa and Affleck\,\cite{Oshikawa1,Affleck,Oshikawa3} discussed this problem on the basis of the model Hamiltonian expressed as
\begin{eqnarray}
\mathcal{H}=\sum_{i} \left[J\bm S_i\cdot \bm S_{i+1}-g{\mu}_{\rm B}HS_i^z-(-1)^ig{\mu}_{\rm B}hS_i^x\right], 
\label{eq:model}
\end{eqnarray}
where $h$ is the staggered field induced by the external field $H$ and is perpendicular to $H$. The staggered field originates from the alternating $\bm g$ tensor and the antisymmetric interaction of the Dzyaloshinsky-Moriya (DM) type with the alternating $\bm D$ vector\,\cite{Moriya}. Thus, the proportional coefficient $c_{\rm s}\,{=}\,h/H$ depends on the field direction. The effect of the staggered field due to the staggered component of the $\bm g$ tensor on spin dynamics was first discussed by Nagata\,\cite{Nagata}. Using the classical spin approach, he succeeded in a qualitative description of the temperature dependence of the resonance field of electron spin resonance (ESR) observed by Oshima {\it et al.}\,\cite{Oshima1} in copper benzoate. Using the field theoretical approach, Oshikawa and Affleck\,\cite{Oshikawa1,Affleck,Oshikawa3} mapped model (\ref{eq:model}) onto the quantum sine-Gordon (SG) model with Lagrangian density 
\begin{eqnarray}
\mathcal{L}=(1/2)\left[({\partial}_{t}{\phi})^2\,{-}\,v_{\rm s}^2({\partial}_{x}{\phi})^2\right]+hC\cos (2{\pi}R{\tilde \phi}),
\label{eq:Lag}
\end{eqnarray}
where $\phi$ is a canonical Bose field, $\tilde \phi$ is the dual field, $R$ is the compactification radius, $v_{\rm s}$ is the spin velocity and $C$ is a coupling constant. The dual field $\tilde \phi$ corresponds to the angle between the transverse component of the spin and the reference direction in a plane perpendicular to the external magnetic field. Oshikawa and Affleck demonstrated that the field-induced gap is given by 
\begin{eqnarray}
{\Delta}(H)\simeq AJ(g{\mu}_{\rm B}h/J)^{2/3}\left[\ln\,(J/g{\mu}_{\rm B}h)\right]^{1/6}, 
\label{eq:gap}
\end{eqnarray}
with $A\,{=}\,1.66$ \cite{Affleck2}. Their results are in good agreement with experimental results for the gap\,\cite{Dender} and the resonance field of low-temperature ESR\,\cite{Oshima2}. After their pioneering work, the elementary excitations and thermodynamic properties in the systems described by model (\ref{eq:model}) and related systems have been actively investigated both theoretically\,\cite{Essler3,Essler2,Lou,Capraro,Wang,Essler1,Zhao,Lou2,Glocke,Orignac,Kuzmenko} and experimentally\,\cite{Asano,Nojiri,Oshikawa2,Kohgi,Feyerherm,Zvyagin,Wolter1,Wolter2,Kenzelmann2,Kenzelmann,Chen}. The substances studied include Yb$_4$As$_3$\,\cite{Oshikawa2,Kohgi}, PM$\cdot$Cu(NO$_3$)$_2\cdot$(H$_2$O)$_2$ (PM=pyrimidine) \cite{Feyerherm,Zvyagin,Wolter1,Wolter2} and CuCl$_2$${\cdot}$2((CD$_3$)$_2$SO)\,\cite{Kenzelmann2,Kenzelmann,Chen} in addition to copper benzoate\,\cite{Asano,Nojiri}.

In the above-mentioned compounds, the exchange constant $J$ ranges from 16 to 36 K and the proportional coefficient is $c_{\rm s}\,{\leq\,}0.08$\,\cite{Nojiri,Zvyagin}. Therefore, the experimental condition has been limited to ${\Delta}(H)\,{<}\,g{\mu}_{\rm B}H$ and $g{\mu}_{\rm B}H/J\,{\geq}\,0.1$. For the comprehensive understanding of the systems described by model (\ref{eq:model}), a new compound having a large interaction constant and large proportional coefficient is necessary. In the previous letter\,\cite{Morisaki}, we reported the results of magnetic susceptibility, specific heat and high-frequency, high-field ESR measurements on the $S\,{=}\,1/2$ AFHC system, KCuGaF$_6$, in external magnetic fields parallel to the $c$ axis. KCuGaF$_6$ can be described by model (\ref{eq:model}) with a large exchange interaction, $J/k_{\rm B}\,{=}\,103$ K. We observed ESR modes identified as soliton resonance and breathers that are characteristic of the quantum SG model. It was found that their resonance conditions can be well explained by the quantum SG field theory\,\cite{Affleck,Essler1} with a rather large staggered field $h\,{\simeq}\,0.17H$.  KCuGaF$_6$ does not order down to 0.46 K, which indicates good one-dimensionality. The upper limit of interchain interaction $J'$ was evaluated as $J'/J\,{<}\,2{\times}10^{-3}$.

ESR is the most powerful tool for detecting $q\,{=}\,0$ excitations with high resolution. Previous ESR measurements on KCuGaF$_6$\,\cite{Morisaki} were performed at $T\,{=}\,1.5$ K for $H\,{\parallel}\,c$ only. In the present work, we carried out high-frequency, high-field ESR experiments for four different field directions at $T\,{=}\,0.5$ K to suppress the thermal effect. As shown below, we observed a variety of excitations, such as soliton resonance, breathers, interbreather transitions and multiple breather excitations, that cannot be explained by the conventional spin wave theory. To investigate the dependence of the staggered field on the external field direction, we also performed magnetic susceptibility measurements by rotating the sample. In this paper, we present the total results of ESR and susceptibility measurements and their analyses based on the quantum SG field theory. \\

\begin{figure}[htbp]
	\begin{center}
		\includegraphics[scale =1.00]{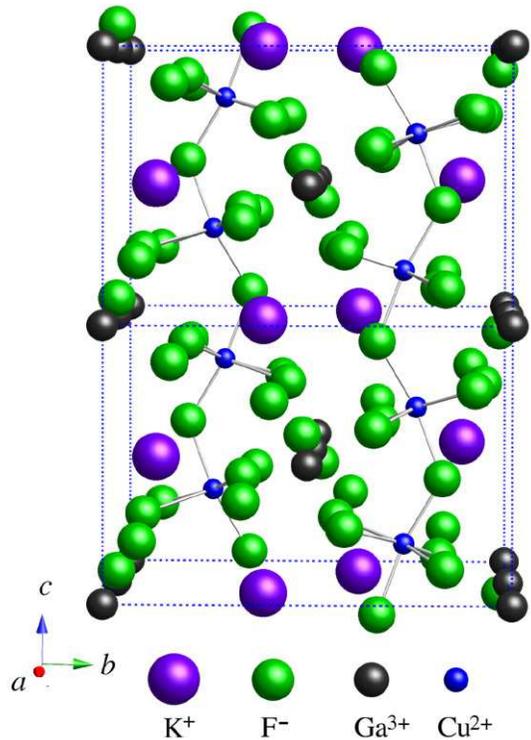}
	\end{center}
	\caption{Perspective view along the $a$ axis of the crystal structure of KCuGaF$_6$. Dotted lines denote the chemical unit cells.}
	\label{fig:KCuGaF6}
\end{figure} 

\section{Crystal Structure and Magnetic Model}
KCuGaF$_6$ crystallizes in a monoclinic structure of space group $P2_1/c$\,\cite{Dahlke}, which is isostructural to KCuCrF$_6$\,\cite{Kissel}. 
The lattice parameters at room temperature are $a\,{=}\,7.2856$\,\AA, $b\,{=}\,9.8951$\,\AA, $c\,{=}\,6.7627$\,\AA, $\beta\,{=}\,93.12^\circ$ and $Z\,{=}\,4$. Figure \ref{fig:KCuGaF6} shows the crystal structure of KCuGaF$_6$. Cu$^{2+}$ and Ga$^{3+}$ ions, both surrounded octahedrally by six F$^-$ ions, form a pyrochlore lattice, in which Cu$^{2+}$ ions with spin-$1/2$ and nonmagnetic Ga$^{3+}$ ions are arranged almost straightly along the $c$ and $a$ axes, respectively. Since the magnetic chain composed of Cu$^{2+}$ ions is separated by other nonmagnetic ions, the exchange interaction between neighboring Cu$^{2+}$ ions should have a one-dimensional (1D) nature. 
CuF$_6$ octahedra are elongated perpendicular to the chain direction parallel to the $c$ axis due to the Jahn-Teller effect. The elongated axes alternate along the chain direction. For this reason, the hole orbitals of Cu$^{2+}$ ions are linked along the chain direction through the $p$ orbitals of F$^-$ ions. The bond angle ${\alpha}$ of the exchange pathway Cu$^{2+}$$\,{-}\,$F$^{-}$$\,{-}\,$Cu$^{2+}$ is ${\alpha}\,{=}\,129^{\circ}$. This large bond angle produces the strong antiferromagnetic exchange interaction $J/k_{\rm B}\,{=}\,103$ K \cite{Morisaki}.  

In KCuGaF$_6$, the local principal axes of octahedra are tilted alternately along the $c$ axis, as shown in Fig. \ref{fig:KCuGaF6}. This leads to the staggered inclination of the principal axes of the ${\bm g}$ tensor. Since there is no inversion symmetry at the middle point of two adjacent spins along the $c$ axis, the DM interaction can exist. Therefore, the magnetic model of KCuGaF$_6$ in external magnetic field $\bm H$ should be expressed as
\begin{eqnarray}
\mathcal{H}=\sum_{i}\Big[J\bm{S}_i{\cdot}\bm{S}_{i+1}-\mu_{\mathrm{B}}(\bm{g}_{i}\bm{H}){\cdot}\bm{S_i}+\bm{D}_{i}{\cdot}[\bm{S}_{i}{\times}\bm{S}_{i+1}]\Big],
\label{eq:model2}
\end{eqnarray}
where the first, second and third terms are the isotropic exchange interaction, the Zeeman term and the DM interaction, respectively. Because of the staggered inclination of the principal axes of the CuF$_6$ octahedra along the $c$ axis, the ${\bm g}$ tensor at the $i$-th spin site is written as 
\begin{eqnarray}
\bm{g}_{i}=\bm{g}_{\rm u}+(-1)^{i}\bm{g}_{\rm s},
\label{eq:g}
\end{eqnarray}
where $\bm{g}_{\rm u}$ is the uniform ${\bm g}$ tensor that is common to all the spin sites and $\bm{g}_{\rm s}$ is the staggered ${\bm g}$ tensor with nondiagonal terms only. Because of the staggered ${\bm g}$ tensor, the staggered magnetic field ${\bm h}^{\rm s}_i\,{=}\,(-1)^i\bm{g}_{\rm s}{\bm H}/g'$ is induced perpendicular to the external magnetic field ${\bm H}$, where $g'$ is the uniform $g$ factor for the staggered field direction. At present, details of the ${\bm g}$ tensor are not clear, because no ESR signal at the X ($\,{\sim}\,9$ GHz) or K ($\,{\sim}\,24$ GHz) band frequency is observed at room temperature owing to large linewidth, which is ascribed to the DM interaction discussed below.

Next we consider the configuration of the $\bm D_i$ vector of the DM interaction in KCuGaF$_6$. The $\bm D_i$ vector is an axial vector given by the nondiagonal components of the angular momenta of adjacent magnetic ions\,\cite{Moriya}. Since there is the $c$ glide plane at ${\pm}\,b/4$, the $ac$ plane component of the $\bm D_i$ vector alternates along the chain direction, but the $b$ component does not. Thus, the $\bm D_i$ vector should be expressed as
\begin{eqnarray}
{\bm D}_i=\left((-1)^iD_x, D_y, (-1)^iD_z\right),
\label{eq:D_i}
\end{eqnarray}  
where the $x$, $y$ and $z$ axes are chosen to be parallel to the $a^*$, $b$ and $c$ axes, respectively. If the $y$ component $D_y$ is negligible, then the $\bm D_i$ vector is expressed as ${\bm D}_i\,{=}\,(-1)^i{\bm D}$. According to the argument by Affleck and Oshikawa\,\cite{Affleck}, the effective staggered field ${\bm h}^{\rm DM}_i$ acting on $\bm{S}_{i}$ is approximated as
\begin{eqnarray}
{\bm h}^{\rm DM}_i \simeq(-1)^i\frac{g}{2g'J}{{\bm H}}\times {\bm D}.
\label{eq:h_DM}
\end{eqnarray} 
Consequently, the total staggered field ${\bm h}_i$ acting on the $i$-th site is given by
\begin{eqnarray}
\bm{h}_i={\bm h}^{\rm s}_i+{\bm h}^{\rm DM}_i\simeq\frac{(-1)^i}{g'}\left[\frac{g}{2J}{\bm{H}}\,{\times}\,{\bm D} +\bm{g}_s\bm{H}\right].
\label{eq:h_st}
\end{eqnarray}
Equation\,(\ref{eq:h_st}) means that the staggered field ${\bm h}_i$ is induced perpendicular to the external magnetic field $\bm H$ and its magnitude is proportional to $H$. Therefore, the model Hamiltonian of the present system can be written as eq.\,(\ref{eq:model}). For simplification, we set $g'\,{=}\,g$ hereafter, and we rewrite $(g'/g)\bm{h}_i$ as $\bm{h}_i$. \\


\section{Experimental Details}
Single KCuGaF$_6$ crystals were grown by the vertical Bridgman method from the melt of an equimolar mixture of KF, CuF$_2$ and GaF$_3$ packed into a Pt tube of 9.6 mm inner diameter and $70{\sim}100$ mm length. One end of the Pt tube was welded and the other end was tightly folded with pliers. The temperature at the center of the furnace was set at 800 $^\circ$C, and the lowering rate was 3 mm/h. The materials were dehydrated by heating in vacuum at about 150$^{\circ}$C for three days. Transparent light-pink crystals with a typical size of $3\,{\times}\,3\,{\times}\,3$ mm$^3$ were obtained. These crystals were identified as KCuGaF$_6$ by X-ray powder diffraction analysis. KCuGaF$_6$ crystals were cleaved along the $(1,1,0)$ plane. The $a$ and $b$ directions were determined by X-ray single-crystal diffraction.

Magnetic susceptibilities were measured using a SQUID magnetometer (Quantum Design MPMS XL) down to 1.8 K. Sample rotation equipment was used to measure the anisotropy of magnetic susceptibility. The high-frequency, high-field ESR measurements were performed in the frequency range of $135\,{-}\,761.6$ GHz using the terahertz electron spin resonance apparatus (TESRA-IMR) \cite{Nojiri} at the Institute for Material Research, Tohoku University. The temperature of the sample was lowered to 0.5 K using liquid ${}^3$He in order to suppress the finite temperature effect. Magnetic field up to 30 T was applied with a multilayer pulse magnet. FIR lasers, backward traveling wave tubes and Gunn oscillators were used as light sources. ESR absorption signals were collected for $H\,{\parallel}\,a$, $H\,{\parallel}\,b$, $H\,{\parallel}\,c$ and $H\,{\perp}\,(1,1,0)$. \\


\section{Results and Discussions}
\subsection{Magnetic susceptibilities}
\begin{figure}[htbp]
  \begin{center}
    \includegraphics[scale =0.50]{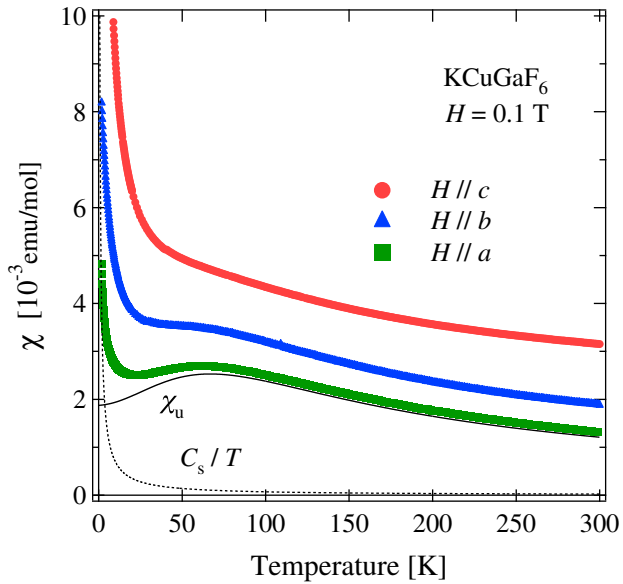}
  \end{center}
  \caption{Temperature dependence of magnetic susceptibilities $\chi$ in KCuGaF$_6$ measured at $H\,{=}\,0.1$ T for $H\,{\parallel}\,a$, $H\,{\parallel}\,b$ and $H\,{\parallel}\,c$. The susceptibilities for $H\,{\parallel}\,b$ and $H\,{\parallel}\,c$ are shifted upward by $5\,{\times}\,10^{-4}$ emu/mol and $2\,{\times}\,10^{-3}$ emu/mol, respectively. Solid and dashed lines denote the uniform component $\chi_{\rm u}$ and the Curie term of the magnetic susceptibility for $H\,{\parallel}\,a$, respectively.}
  \label{fig:susceptibility}
\end{figure}

\begin{figure}[htbp]
  \begin{center}
    \includegraphics[scale =0.45]{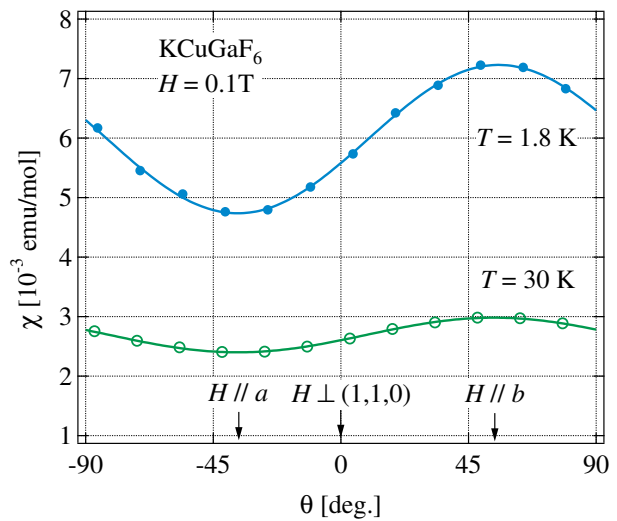}
  \end{center}
  \caption{Dependence of magnetic susceptibility on field direction in the $ab$ plane. Susceptibility data were collected at $T\,{=}\,1.8$ and 30 K. Solid lines are fits by sine curves with a period of 180$^{\circ}$. $\theta$ denotes the angle between the external field and the normal of the cleavage $(1,1,0)$ plane.}
  \label{fig:rotator}
\end{figure}

First we show the temperature dependence of the magnetic susceptibilities $\chi$ measured at $H\,{=}\,0.1$ T for $H\,{\parallel}\,a$,  $H\,{\parallel}\,b$ and $H\,{\parallel}\,c$ in Fig.\ \ref{fig:susceptibility}. With decreasing temperature, the magnetic susceptibilities increase rapidly below 30 K, obeying the Curie law. The Curie constant $C_{\rm st}$ depends strongly on the field direction and is independent of the specimen. Thus, the Curie term is intrinsic to the present system. $C_{\rm st}$ is the largest for $H\,{\parallel}\,c$ and the smallest for $H\,{\parallel}\,a$. For $H$ applied in the $ab$ plane, $C_{\rm st}$ becomes the largest for $H\,{\parallel}\,b$, as shown in Fig\ \ref{fig:rotator}.
The susceptibilities for $H\,{\parallel}\,a$ and $H\,{\parallel}\,b$ exhibit broad peaks at $T_{\rm max}\,{\sim}\,70$ K, which are characteristic of AFHC. The broad peak is clearly observed for $H\,{\parallel}\,a$, while for $H\,{\parallel}\,c$, it is completely hidden by the large Curie term.

The magnetic susceptibility for model (\ref{eq:model}) was calculated by Affleck and Oshikawa\,\cite{Affleck} as 
\begin{eqnarray}
{\chi}=-\,\frac{d^2F}{dH^2}={\chi}_{\rm u} + c_{\rm s}^2{\chi}_{\rm s}\,,
\label{eq:chi_1}
\end{eqnarray}
where $F$ is the free energy, $\chi_{\rm u}$ is the uniform magnetic susceptibility for $S\,{=}\,1/2$ AFHC without the staggered field, which is shown in Refs.\,\cite{Bonner,Eggert,Johnston}, and $\chi_{\rm s}$ is the staggered susceptibility given by
\begin{eqnarray}
\chi_{\rm s}(T)\simeq 0.278\left(\frac{N_{\mathrm{A}}g^2\mu_{\mathrm{B}}^2}{4k_{\mathrm{B}}T}\right)\left \{ \ln \left(\frac{J}{k_{\mathrm{B}}T}\right)\right \}^{1/2}.
\label{eq:chi_s}
\end{eqnarray}
The magnetic susceptibility $\chi$ given by eq.\,(\ref{eq:chi_1}) can be obtained from the magnetic force measurement by the Faraday method. The magnetic force acting on a sample that is placed in a magnetic field with a field gradient arises not only from the uniform magnetization $M$ induced along the external field $H$ but also from the staggered magnetization $M_{\rm s}$ induced perpendicular to $H$, because the staggered field $h$ increases with increasing external magnetic field, which lowers the Zeeman energy $({-}\,M_{\rm s}h)$. In the present experiments, however, we measure the uniform magnetization using only the detection coil, in which the magnetized sample moves. The DM interaction contributes to the the uniform magnetization, but the Zeeman term due to the staggered $\bm g$ tensor does not, because the latter interaction does not give rise to the canting of spins toward the external field direction. Thus, the magnetic susceptibility defined as ${\chi}\,{=}\,dM/dH$ should be expressed as
\begin{eqnarray}
{\chi}={\chi}_{\rm u} + \frac{D_{\perp}}{2J}c_{\rm s}{\chi}_{\rm s}\cos{\beta}\,,
\label{eq:chi_2}
\end{eqnarray}
where $D_{\perp}\,{=}\,|{\bm{H}}\,{\times}\,{\bm D}|/H$ and $\beta$ is the angle between ${\bm h}^{\rm DM}_i$ and ${\bm h}_i$. When the DM interaction is absent, the Curie term vanishes.

The staggered susceptibility $\chi_{\rm s}$ obeys the Curie law for $T\,{\ll}\,J/k_{\rm B}$ where the subleading logarithmic term of $\chi_{\rm s}$ is almost constant. Thus, we can deduce that the Curie term in the magnetic susceptibility of KCuGaF$_6$ arises from the temperature dependence of the staggered susceptibility. However, with increasing temperature, $\chi_{\rm s}$ decreases more rapidly due to the subleading term that becomes zero at $T\,{=}\,J/k_{\rm B}$. Therefore, the Curie term observed in the magnetic susceptibility of KCuGaF$_6$ is not successfully describable by $\chi_{\rm s}$ of eq.\,(\ref{eq:chi_s}). The subleading term of $\chi_{\rm s}$ is valid only for $T\,{\ll}\,J/k_{\rm B}$. At present, there is no analytical result on the subleading term that is applicable for $T\,{\sim}\,J/k_{\rm B}$. 
 
The magnetic susceptibility $\chi$ is best described as  
\begin{eqnarray}
{\chi}=\chi_{\rm u}+\frac{C_{\rm st}}{T}\,.
\label{eq:sus}
\end{eqnarray}
The Curie constant $C_{\rm st}$ should be proportional to $(D_{\perp}/J)c_{\rm s}\cos{\beta}$. Solid and dashed lines in Fig.\ \ref{fig:susceptibility} denote the uniform susceptibility $\chi_{\rm u}$ and the Curie term obtained by fitting eq.\,(\ref{eq:sus}) to the susceptibility for $H\,{\parallel}\,a$, respectively. From the uniform susceptibilities obtained by the fits for three field directions for $H\,{\perp}\,c$, we obtain $J/k_{\rm B}\,{=}\,103\,{\pm}\,2$\,K. This exchange constant coincides with that obtained from the low-temperature specific heat at zero magnetic field\,\cite{Morisaki}, which is described as $C(T)\,{=}\,{\gamma}T\,{+}\,bT^3$ with ${\gamma}\,{=}\,2Rk_{\rm B}/(3J)$. The Curie constants $C_{\rm st}$ obtained by fitting eq.\,(\ref{eq:sus}) to the susceptibilities for four different field directions are summarized in Table\ \ref{table:cs}.

\begin{table}
\caption{Curie constant $C_{\rm st}$ and proportional coefficient $c_{\rm s}\,{=}\,h/H$ obtained for $H\,{\parallel}\,c$, $H\,{\parallel}\,b$, $H\,{\perp}\,(1,1,0)$ and $H\,{\parallel}\,a$.}
\label{table:cs}
\begin{ruledtabular}
\begin{tabular}{lcccc} 
& $H\,{\parallel}\,c$ & $H\,{\parallel}\,b$ & $H\,{\perp}\,(1,1,0)$ & $H\,{\parallel}\,a$ \\ \hline
 $C_{\rm st}$ [emu$\cdot$K/mol]  & 0.051 & 0.014 & 0.012 & 0.006 \\
$c_{\rm s}$ & 0.178 & 0.160 & 0.056 & 0.031 \\
\end{tabular}
\end{ruledtabular}
\end{table}

Figure\ \ref{fig:rotator} shows the dependence of magnetic susceptibility on the field direction measured at $T\,{=}\,1.8$ and 30 K. The external magnetic field of 0.1 T was applied in the $ab$ plane. The angle $\theta$ is the angle between the external field and the normal of the cleavage $(1,1,0)$ plane. At room temperature, the magnetic susceptibility exhibits a maximum for $H\,{\parallel}\,b$ and a minimum for $H\,{\parallel}\,a$. This susceptibility behavior is due to the anisotropy of the $g$ factor. With decreasing temperature, the anisotropy of the magnetic susceptibility increases rapidly below 30 K. This is attributed to the anisotropy of the coefficient $(D_{\perp}/J)c_{\rm s}\cos{\beta}$ and the temperature dependence of the staggered susceptibility ${\chi}_{\rm s}$ in eq.\,(\ref{eq:chi_2}). At low temperatures, the maximum and minimum susceptibilities occur for $H\,{\parallel}\,b$ and $H\,{\parallel}\,a$, respectively, which implies that the staggered field becomes maximum for $H\,{\parallel}\,b$ and minimum for $H\,{\parallel}\,a$ in the $ab$ plane. From the anisotropy of the magnetic susceptibility, it is expected that in KCuGaF$_6$, the staggered field $h$ depends strongly on the direction of the external magnetic field $H$, and that the staggered field is the largest for $H\,{\parallel}\,c$ and the smallest for $H\,{\parallel}\,a$. \\

\subsection{Electron spin resonance (ESR)}

\begin{figure}[htbp]
\begin{center}
 \includegraphics[scale =0.65]{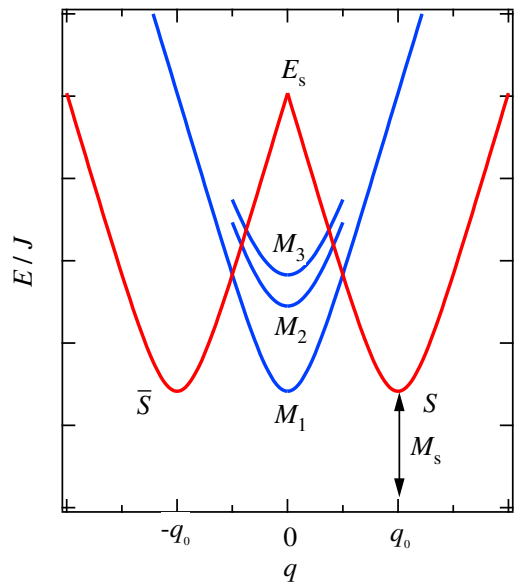}
\end{center}
\caption{Illustration of low-energy excitations of model (\ref{eq:model}) around $q\,{=}\,0$. Soliton, antisoliton, soliton resonance and three breathers are labeled as $S$, $\bar S$, $E_{\rm s}$ and $M_1\,{\sim}\,M_3$, respectively.} 
\label{fig:excitations}
\end{figure}

Figure \ref{fig:excitations} illustrates low-energy excitations around $q\,{=}\,0$. Because of the staggered field $h$ induced by the external magnetic field, the gapless excitations at $q\,{=}\,0$ and ${\pm}\,q_0$ for $h\,{=}\,0$ have finite gaps, where $q_0$ is an incommensurate wave vector given by $q_0\,{=}\,2{\pi}m(H)$ in the absence of the staggered field. In the quantum SG model, low-energy elementary excitations are composed of solitons, antisolitons and their bound states called breathers. The soliton mass $M_{\rm s}$ corresponds to the excitation energy at $q\,{=}\,{\pm}\,q_0$ and ${\pi}\,{\pm}\,q_0$. 
Essler {\it et al.}\,\cite{Essler1} calculated the soliton mass that is applicable in a wide magnetic field range for $0\,{<}\,H\,{<}\,H_{\rm s}$, where $H_{\rm s}$ is the saturation field given by $H_{\rm s}\,{=}\,2J/g{\mu}_{\rm B}$. Their result is expressed as
\begin{eqnarray}
\frac{M_{\rm s}}{J}\,{=}\,\frac{2v}{\sqrt{\pi}}\frac{\Gamma \left(\displaystyle\frac{\xi}{2}\right)}{\Gamma \left(\displaystyle\frac{1+\xi}{2}\right)}\left[\frac{\Gamma \left(\displaystyle\frac{1}{1+\xi}\right)}{\Gamma \left(\displaystyle\frac{\xi}{1+\xi}\right)} \frac{c{\pi}g{\mu}_{\rm B}H}{2Jv}c_{\rm s}\right]^{(1+\xi)/2},\ \ 
\label{eq:solitonmass}
\end{eqnarray}
where $v$ is the dimensionless spin velocity, $\xi$ is a parameter given by ${\xi}\,{=}\,[2/({\pi}R^2)-1]^{-1}$ and $c$ is a parameter depending on magnetic field. The field dependences of these parameters are shown in the literature \cite{Affleck,Essler1,Hikihara}. For $H \rightarrow$ 0, $v\,{\rightarrow}\,{\pi}/2$, $\xi\,{\rightarrow}\,1/3$ and $c\,{\rightarrow}\,1/2$. 

The breathers corresponding to the excitations at $q\,{=}\,0$ and ${\pi}$ have hierarchical structures labeled by integer $n\,({=}\,1,\,2,\,\cdots)$. The mass of the $n$-th breather is determined by the soliton mass $M_{\rm s}$ and parameter $\xi$ as
\begin{eqnarray}
M_n=2M_{\rm s} {\sin}\,\left(\frac{n{\pi}{\xi}}{2}\right).
\label{eq:breather}
\end{eqnarray}
The number of breathers is limited by $n\,{\leq}\,[{\xi}^{-1}]$\,\cite{Affleck}. In our experimental field range, $g{\mu}_{\rm B}H/J\,{<}\,0.5$, breathers up to the third order can be observed.

\begin{figure}[htbp]
\begin{center}
\includegraphics[keepaspectratio=true,width=88mm]{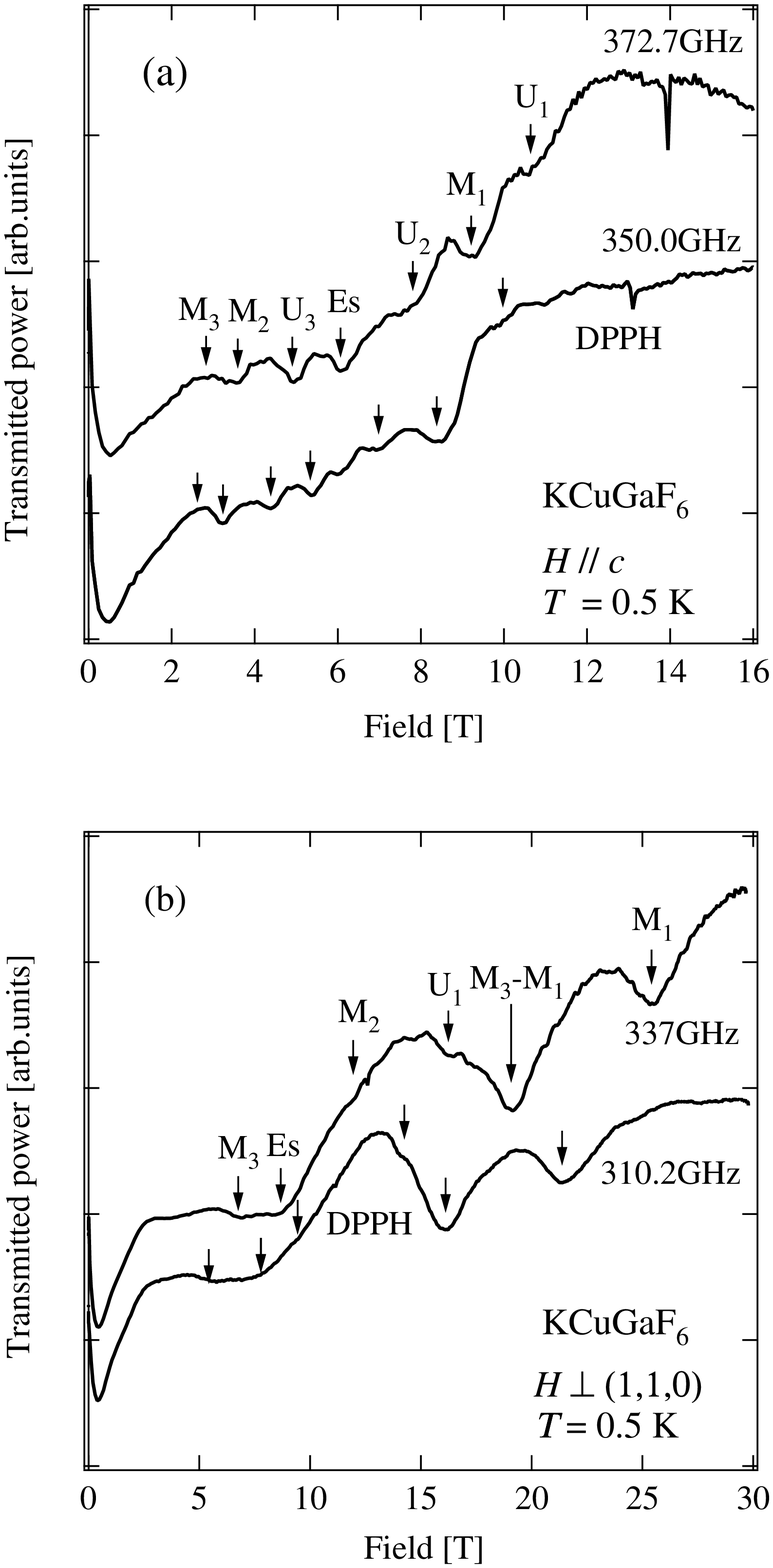}
\end{center}
\caption{Examples of ESR absorption spectra in KCuGaF$_6$ measured at $T\,{=}\,0.5$ K for (a) $H\,{\parallel}\,c$ and (b) $H\,{\perp}\,(1,1,0)$. Arrows labeled $E_{\rm s}$, $M_n$ and $M_n-M_{n'}$ denote the fields for soliton resonance, $n$-th breather and interbreather transition, respectively. Resonances labeled $U_n$ are those whose origins are unclear. Sharp absorption labeled DPPH is the marker of $g\,{=}\,2.00$.} 
\label{fig:ESR_spectra}
\end{figure}

\begin{figure*}[htbp]
\begin{center}
    \includegraphics[keepaspectratio=true,width=181mm]{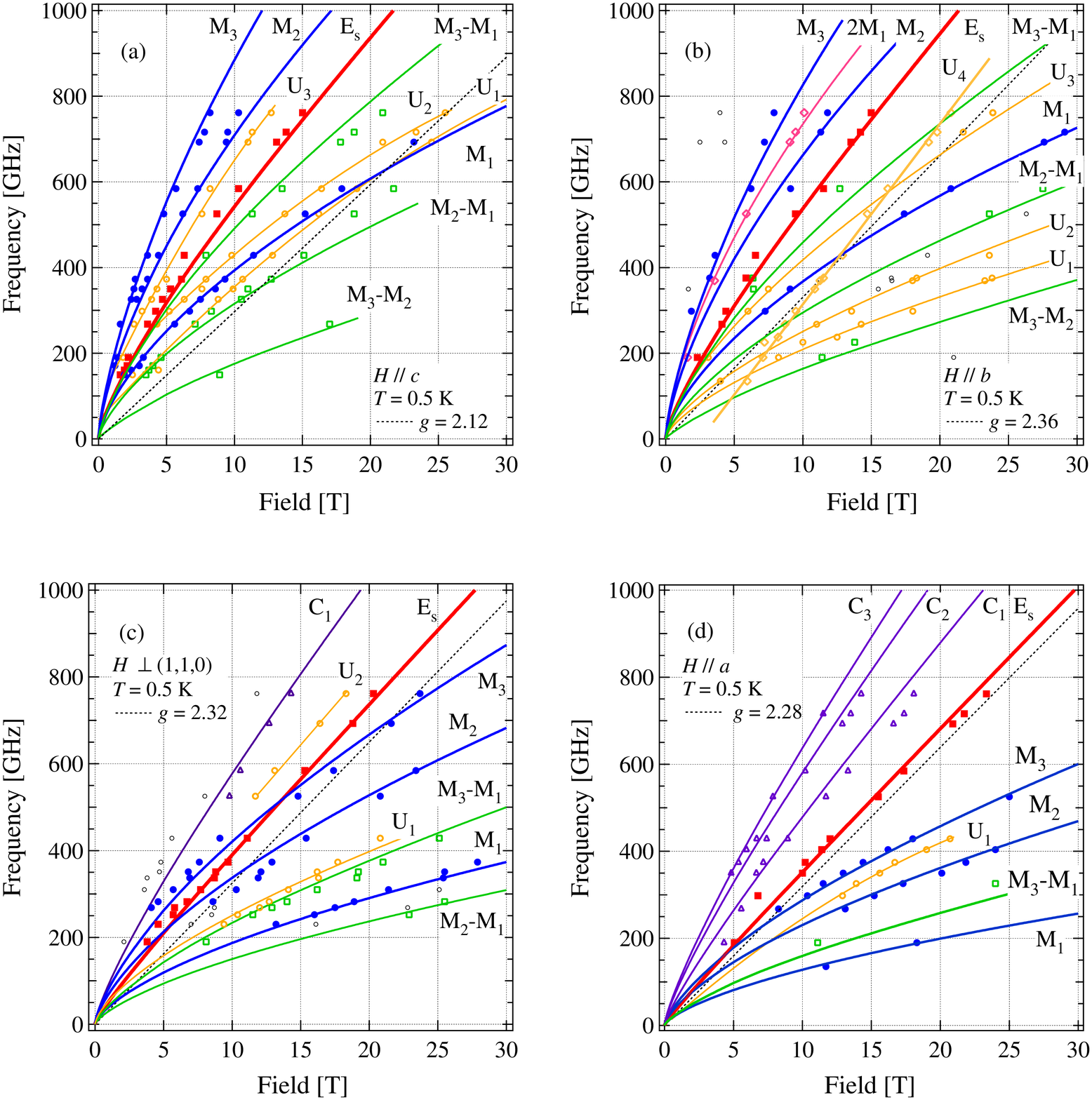}
  \end{center}
   \caption{Frequency vs field diagrams of ESR modes in KCuGaF$_6$ measured at $T\,{=}\,0.5$ K for (a) $H\,{\parallel}\,c$, (b) $H\,{\parallel}\,b$, (c) $H\,{\perp}\,(1,1,0)$ and (d) $H\,{\parallel}\,a$. Symbols denote experimental resonance fields. Thick and thin solid lines and dashed lines labeled as $E_{\rm s}$, $M_n$, $M_n-M_{n'}$ and $2M_1$ are resonance conditions calculated from eqs.\,(\ref{eq:solitonmass}), (\ref{eq:breather}) and (\ref{eq:soliton_resonance}) with $J/k_{\rm B}\,{=}\,103$ K and the proportional coefficient $c_{\rm s}$ shown in Table\,\ref{table:cs}. }
  \label{fig:ESR_diagram2}
\end{figure*}

To observe elementary excitations in KCuGaF$_6$, we performed high-frequency ESR measurements combined with pulsed high magnetic field at 0.5 K. Since KCuGaF$_6$ has a large exchange interaction of $J/k_{\rm B}\,{=}\,103$ K, we are able to observe elementary excitations in the relatively low-field region over a wide energy range as compared with copper benzoate\,\cite{Asano,Nojiri} and PM$\cdot$Cu(NO$_3$)$_2\cdot$(H$_2$O)$_2$\,\cite{Zvyagin}. In ESR measurements, we can observe only $q\,{=}\,0$ excitations. Therefore, the breathers can be observed by ESR, whereas the soliton and antisoliton corresponding to the excitations at $q\,{=}\,{\pm}\,q_0$ cannot be observed directly. Instead, we can observe a soliton resonance labeled $E_{\rm s}$ in Fig.\,\ref{fig:excitations}, which corresponds to the excitation energy at $q=0$ on the excitation branch connected to the soliton and antisoliton at $q\,{=}\,{\pm}\,q_0$\,\cite{Affleck,Zvyagin}. The condition of soliton resonance is given by 
\begin{equation}
E_{\rm s} \simeq \sqrt{M_{\rm s}^2+(g\mu_{\mathrm{B}}H)^2}.
\label{eq:soliton_resonance}
\end{equation}

Figure \ref{fig:ESR_spectra} shows examples of ESR spectra obtained at $T\,{=}\,0.5$ K for $H\,{\parallel}\,c$ and $H\,{\perp}\,(1,1,0)$. Absorption signals observed upon sweeping field both up and down were determined as intrinsic resonance signals. The resonance data are summarized in Fig.\ \ref{fig:ESR_diagram2}. As many as about ten resonance modes were observed for four different field directions. This result cannot be explained in terms of the linear spin wave theory, which yields only two excitation modes. 

Resonance modes that were assigned as soliton resonance and breathers are labeled as $E_{\rm s}$ and $M_n$ ($n\,{=}\,1\,{\sim}\,3$), respectively, in Figs.\ \ref{fig:ESR_spectra} and \ref{fig:ESR_diagram2}. Thick solid lines in Fig.\ \ref{fig:ESR_diagram2} denote their resonance conditions calculated from eqs.\,(\ref{eq:solitonmass}), (\ref{eq:breather}) and (\ref{eq:soliton_resonance}) with exchange constant $J/k_{\rm B}\,{=}\,103$ K and proportional coefficient $c_{\rm s}\,{=}\,h/H$ shown in Table\ \ref{table:cs}, where the error of $c_{\rm s}$ is ${\pm}\,0.005$. The soliton resonance and the breathers up to the third order are the main excitations that are predicted by the quantum SG field theory. In KCuGaF$_6$, all of these excitations were clearly observed for four different field directions. As shown in Fig.\ \ref{fig:ESR_diagram2}, the experimental results are successfully described by the quantum SG field theory with only adjustable parameter $c_{\rm s}$. In these calculations, we used the $g$ factor $g\,{=}\,2.32$ for $H\,{\perp}\,(1,1,0)$, which was determined by the present ESR measurement at $T\,{\sim}\,60$ K. The $g$ factors used for $H\,{\parallel}\,a$, $H\,{\parallel}\,b$ and $H\,{\parallel}\,c$ were determined from the uniform magnetic susceptibilities ${\chi}_{\rm u}$ at room temperature as $g\,{=}\,2.28$, 2.36 and 2.12, respectively, assuming that ${\chi}_{\rm u}/g^2$ is constant. As shown in Table\ \ref{table:cs}, the proportional coefficient $c_{\rm s}$ varies from 0.031 to 0.178.  For $H\,{\parallel}\,c$ and $H\,{\parallel}\,b$, $c_{\rm s}\,{=}\,0.178$ and 0.160, respectively. Because of the large $c_{\rm s}$, the soliton mass $M_{\rm s}$ is larger than $g{\mu}_{\rm B}H$ in the present magnetic field range. Such a large proportional coefficient has not been observed in other SG systems. On the other hand, for $H\,{\parallel}\,a$, $M_{\rm s}$ is smaller than $g{\mu}_{\rm B}H$ because of small $c_{\rm s}$ ({=}\,0.031).

For $H\,{\leq}\,M_{\rm s}\,{\ll}\,J$, the soliton mass $M_{\rm s}$ is given by $\Delta$ in eq.\,(\ref{eq:gap})\,\cite{Affleck}. This condition is satisfied for $H\,{\parallel}\,c$ and $H\,{\leq}\,20$ T. When we use eq.\,(\ref{eq:gap}) instead of eq.\,(\ref{eq:solitonmass}) to calculate the resonance conditions for $H\,{\parallel}\,c$, we obtain $c_{\rm s}\,{=}\,0.25$, which is 1.5 times as large as $c_{\rm s}$ obtained using eq.\,(\ref{eq:solitonmass}).

The intensities of the soliton resonance $E_{\rm s}$ and the breathers $M_n$ are of the same order. 
ESR is caused by the oscillating magnetic field $H_1$ of the light. The soliton resonance occurs when $H_1$ is perpendicular to the external field $H$, while breathers are excited when $H_1$ is parallel to $H$. In the classical picture of antiferromagnetic resonance, the motion of the total magnetization corresponding to the soliton resonance is the precession around the external field $H$, and that for the breather is the oscillation parallel to $H$. Since the diameter of the light pipe is larger than the wavelength of the light used in the present experiments, the light propagates in a light pipe parallel to the external magnetic field with several propagation modes. Consequently, the oscillating magnetic field has components both parallel and perpendicular to the external field. Thus we can observe both soliton resonance and breathers.

There are resonance modes labeled $M_2\,{-}\,M_1$, $M_3\,{-}\,M_1$ and $M_3\,{-}\,M_2$, as shown in Fig.\ \ref{fig:ESR_diagram2}. Their excitation energies are equal to the energy differences between two of three breathers, as shown by thin dashed lines in Fig.\ \ref{fig:ESR_diagram2}. Within the framework of the quantum SG field theory, there is no excitation from the ground state that has energy $M_n\,{-}\,M_{n'}$. Hence, it is natural to consider that these excitations are transitions between breathers. As shown in Fig.\ \ref{fig:ESR_spectra}(b), the intensities of the interbreather transitions and breathers are of the same order even at 0.5 K, which is much lower than the breather mass. Because the width of the pulsed magnetic field is about 10 msec in the present ESR measurements, the splitting of the excitation levels by applied field occurs under an almost adiabatic condition, maintaining the population at zero magnetic field. Thus, the interbreather transitions are observable even at 0.5 K.

For $H\,{\parallel}\,b$, a resonance mode labeled $2M_1$ is observed. Because its excitation energy is just twice the mass of the first breather $M_1$, the $2M_1$ mode is the simultaneous excitation of two first breathers. Such two-breather resonance was observed for the first time in the present system. The energy of two-soliton excitation $2M_{\rm s}$ almost coincides with the energy of the third breather. Therefore, it seems difficult to observe the two-soliton excitation, if it exists. 
There are additional weak resonance modes labeled $C_n$ with $n\,{=}\,1, 2$ and 3 that have high excitation energies. These modes are considered to be the multiple excitations of the soliton resonance and the $n$-th breather, because its energy corresponds to $E_{\rm s}\,{+}\,M_n$.  

Resonance modes $U_1$ to $U_4$, denoted by thin solid lines in Fig.\,\ref{fig:ESR_diagram2}, are those whose origins are unexplainable. These unknown modes are labeled in increasing order of excitation energy at high fields above 20 T. Note that for $H\,{\parallel}\,b$, an unknown linear mode $U_4$ is the most intense. Because the resonance conditions of the three unknown modes ($U_1\,{\sim}\,U_3$) for $H\,{\parallel}\,c$ observed in the present measurements are almost the same as those observed in the previous measurements using different specimen, these modes seems intrinsic to KCuGaF$_6$. These unknown modes were also observed in another quantum SG system, PM$\cdot$Cu(NO$_3$)$_2\cdot$(H$_2$O)$_2$\,\cite{Zvyagin}. The origins of these unknown modes are an open question. 

As shown in Table\,\ref{table:cs}, $c_{\rm s}^2$ is not necessarily proportional to the Curie constant $C_{\rm st}$ in eq.\,(\ref{eq:sus}). It is considered that the Curie constant $C_{\rm st}$ is proportional to $(D_{\perp}/J)c_{\rm s}\cos{\beta}$ in eq.\,(\ref{eq:chi_2}). Hence, the Curie term in the magnetic susceptibility vanishes when the DM interaction is absent, even if the transverse staggered field is induced due to the staggered component of the $\bm g$ tensor. Although $C_{\rm st}$ for $H\,{\parallel}\,b$ is about a quarter of that for $H\,{\parallel}\,c$, the proportional coefficients $c_{\rm s}$ for these two field directions are approximately equal. This indicates that $D_{\perp}\cos{\beta}$ for $H\,{\parallel}\,c$ is about four times as large as that for $H\,{\parallel}\,b$. Because the $\bm D$ vector is assumed to be parallel to the $ac$ plane, as discussed in Section II, the magnitude of $D_{\perp}$ for $H\,{\parallel}\,b$ should be larger than that for $H\,{\parallel}\,c$. Thus, $\cos{\beta}$ for $H\,{\parallel}\,b$ is considerably smaller than that for $H\,{\parallel}\,c$. This implies that the staggered fields due to the DM interaction and the staggered ${\bm g}$ tensor, ${\bm h}^{\rm DM}_i$ and ${\bm h}^{\rm s}_i$, respectively, are roughly parallel for $H\,{\parallel}\,c$, while for $H\,{\parallel}\,b$, they are roughly orthogonal.  \\

\section{Conclusions}
In conclusion, we have presented the results of the magnetic susceptibility and the high-frequency, high-field ESR measurements on $S\,{=}\,1/2$ AFHC KCuGaF$_6$ with the large exchange interaction $J/k_{\rm B}\,{=}\,103$ K. In KCuGaF$_6$, the staggered magnetic field $\bm h_i$ is induced perpendicular to the external magnetic field $\bm H$ owing to the DM interaction with alternating ${\bm D}$ vectors and the staggered $\bm g$ tensor. Thus, the present system can be represented by the quantum SG model in a magnetic field. 
In ESR measurement, we observed the soliton resonance $E_{\rm s}$ and the breathers $M_n$ up to the third order, which are main elementary excitations characteristic of the quantum SG model. We also observed the interbreather transitions $M_n\,{-}\,M_{n'}$, the multiple excitations of the soliton resonance and the breather $E_{\rm s}\,{+}\,M_n$ and two-breather excitation $2M_1$. As shown in Fig.\ \ref{fig:ESR_diagram2}, their resonance conditions for four different field directions were beautifully described by the quantum SG field theory with only the adjustable parameter $c_{\rm s}\,{=}\,h/H$ listed in Table \ref{table:cs}. The proportional coefficient $c_{\rm s}$ varies widely, $0.031\,{\leq}\,c_{\rm s}\,{\leq}\,0.178$, depending on the field direction. KCuGaF$_6$ differs from other quantum SG systems in its large exchange interaction and wide range of the proportional coefficient.
It was shown from the present measurements that the Curie term $C_{\rm st}/T$ in the magnetic susceptibility is not necessarily proportional to $c_{\rm s}^2$. This is because the Curie term arises from the DM interaction but not from the staggered $\bm g$ tensor.  \\


\begin{acknowledgments}
We express our sincere thanks to R. Morisaki for her contribution to the early stage of the present study and to M. Oshikawa and T. Hikihara for fruitful discussions and comments. This work was supported by a Grant-in-Aid for Scientific Research (A) from the Japan Society for the Promotion of Science, and by a Global COE Program ``Nanoscience and Quantum Physics'' at Tokyo Tech and a Grant-in-Aid for Scientific Research on Priority Areas ``High Field Spin Science in 100 T'', both funded by the Japanese Ministry of Education, Culture, Sports, Science and Technology. 
\end{acknowledgments}



\begin{thebibliography}{99}
\bibitem{Affleck2} The coefficient $A$ is given as $A\,{=}\,1.85$ in Ref.\,\cite{Oshikawa1}. In Ref.\,\cite{Affleck} and its Errata, the value of $A$ is corrected as $A\,{=}\,1.78\,{\times}\,(2/3)^{1/6}\,{=}\,1.66$.
\end{thebibliography}
\end{document}